\documentclass{ssdbm}

\usepackage[update,prepend]{epstopdf} 
\usepackage{graphicx}
\usepackage{color}
\usepackage{listings}
\usepackage{aas_macros}
\usepackage{float}
\usepackage[labelfont=bf, textfont=bf]{caption}
\usepackage[sort&compress]{natbib}

\newfloat{copyrightbox}{thp}{lop}

\newfloat{query}{thp}{lop}
\floatname{query}{Query}

\newfloat{listing}{t}{lop}
\floatname{listing}{Listing}

\definecolor{light-gray}{gray}{0.95}
\definecolor{light-blue}{rgb}{0.95,0.95,1.0}
\definecolor{dark-green}{rgb}{0.0,0.5,0.0}

\def\sharedaffiliation{
\end{tabular}
\begin{tabular}{c}
}

\def\csharp{C\nolinebreak\hspace{-.05em}\raisebox{.3ex}{\scriptsize\bf \#\ }}

\begin{document}

\title{Graywulf: A platform for federated scientific databases and services}

\numberofauthors{5} 
%
\author{
%
%
\alignauthor L\'aszl\'o Dobos \\
\alignauthor Istv\'an Csabai \\
\sharedaffiliation
	\affaddr{E\"otv\"os Lor\'and University, Department of Physics of Complex Systems}   \\
	\affaddr{P\'azm\'any P\'eter s\'et\'any 1/A., 1117 Budapest, Hungary } \\
	\\
\and
\alignauthor Alexander S. Szalay \\
\alignauthor Tam\'as Budav\'ari \\
\alignauthor Nolan Li \\
\and  
\sharedaffiliation
	\affaddr{The Johns Hopkins University, Department of Physics \& Astronomy}   \\
	\affaddr{3800 San Martin Drive, Baltimore, MD 21218, USA }
}

\date{30 July 1999}

\maketitle
\begin{abstract}

Many fields of science rely on relational database management systems to analyze, publish and share data. Since RDBMS are originally designed for, and their development directions are primarily driven by, business use cases they often lack features very important for scientific applications. Horizontal scalability is probably the most important missing feature which makes it challenging to adapt traditional relational database systems to the ever growing data sizes. Due to the limited support of array data types and metadata management, successful application of RDBMS in science usually requires the development of custom extensions. While some of these extensions are specific to the field of science, the majority of them could easily be generalized and reused in other disciplines. With the Graywulf project we intend to target several goals. We are building a generic platform that offers reusable components for efficient storage, transformation, statistical analysis and presentation of scientific data stored in Microsoft SQL Server. Graywulf also addresses the distributed computational issues arising from current RDBMS technologies. The current version supports load balancing of simple queries and parallel execution of partitioned queries over a set of mirrored databases. Uniform user access to the data is provided through a web based query interface and a data surface for software clients. Queries are formulated in a slightly modified syntax of SQL that offers a transparent view of the distributed data. The software library consists of several components that can be reused to develop complex scientific data warehouses: a system registry, administration tools to manage entire database server clusters, a sophisticated workflow execution framework, and a SQL parser library.

\end{abstract}




\section{Introduction}

In this paper we outline our Graywulf\footnote{In honor of Jim Gray, the name of our database server cluster software comes from the combination of his last name and from the name of Beowulf HPC clusters.} project to build an extensive and integrated software library and a collection of user interfaces that offer generic solutions to many problems of scientific data processing using database server farms. The potential of RDBMS in scientific computing is already well-explored and shortcomings are mostly identified \cite{DBLP:dblp_conf/icde/StonebrakerAKS12}. The lack of an array data type and transparent horizontal scaling are probably the most important issues. These issues are directly addressed by certain database systems \cite{DBLP:dblp_journals/debu/IdreosGNMMK12, DBLP:dblp_conf/ssdbm/StonebrakerBPR11}, but we believe that building on the basis of the mainstream, business-oriented database products also has its potential in scientific computing.

The paper is structured as follows. After a more detailed introduction, in Section~\ref{sec:rdbms} we discuss the opportunities and challenges of using relational database management systems for scientific data processing, as well as our solution to the identified problems. Section~\ref{sec:query} is about how distributed query execution is implemented in Graywulf, and what new features we are planning to add in the future. Section~\ref{sec:transformations} summarizes our concept of a scientific data exchange that supports integration of data models and file formats into a unified framework. Problems and solutions related to database server cluster management are detailed in Section~\ref{sec:cluster}. Section~\ref{sec:jobs} describes the job system developed for Graywulf and Section~\ref{sec:ui} is about user interfaces and data visualization inside the framework.

\subsection{Relational databases as platform for scientific data analysis}

While originally not designed for scientific applications, during the last decade relational database management systems (RDBMS) have become an essential tool for analyzing, publishing and sharing data in many fields of natural and engineering sciences \cite{2001cs.......11015S, 2006ASPC..351..212L, DBLP:dblp_conf/minenet/MatrayCHSDV07, DBLP:dblp_journals/ijsnet/TerzisMCSSGOLGB10, 2008JTurb...9...31L}. Scientific data processing tasks are typical data warehouse problems in many senses. Database sizes are in the multi-terabyte range, ad-hoc queries are usual and to be executed by a batch system, data are much more often read than written and read operations have to be optimized for sequential access, and there is no need for transaction processing.

Lately, the way of interacting with scientific databases has changed significantly. Traditionally, software engineers built databases that scientists could access via client tools, which were simple web forms in most cases. This approach limited the possibilities of scripting data access, data reduction and analysis steps, whereas scriptability is essential for many fields where complex data processing tasks have to be repeated numerous times. Allowing user access to databases directly via SQL queries solved this problem but long-running queries and large result sets made it necessary to go further and build user interfaces that allowed much more than submitting simple queries and downloading results synchronously. Szalay et al.  developed the concept of \textit{CasJobs}\cite{DBLP:journals/corr/abs-cs-0502072, 50716924, 6071419} for \textit{SkyServer}\cite{2001cs.......11015S}. In CasJobs, registered users automatically get a small database, called \textit{MyDB}, that they can use as a sandbox for data processing. Queries are submitted on a web interface and results are written into the MyDB. Tables in the MyDB can be used in other queries, shared with others or downloaded. MyDBs are co-located with the large databases containing the science archives, thus join queries between user data and the archives can be executed. User queries are scheduled by a queue system to prevent the congestion of servers by concurrently submitted expensive queries.

Scientists have learned how to translate their data mining problems into SQL~queries and numerous extensions have been developed to support the special needs of researchers. For instance, Microsoft SQL~Server has been extended with an Array Library \cite{DBLP:dblp_conf/edbt/DobosSBBCTMTJ11, 2012ASPC..461..323D} that helps store multi-dimensional arrays in variables and tables, offers a set of flexible array transformation functions including arbitrary slicing, and provides wrapper functions around the LAPACK and FFTW libraries to call high performance mathematical functions directly from SQL. Other extensions, like the Spherical Library \cite{2007ASPC..376..559B} with Hierarchical Triangular Mesh \cite{2010PASP..122.1375B}, or various 3D and kD search libraries \cite{DBLP:dblp_conf/ssdbm/LemsonBS11, DBLP:dblp_conf/cidr/CsabaiTDJHPBS07, 2007AN....328..852C}, provide support for spatial queries on the surface of the sphere or in multi-dimensional space at a resolution required by astronomy. Functions of these libraries can be accessed via user-defined functions (UDFs). Because using UDFs to manipulate complex data structures, especially arrays, is rather cumbersome, a good strategy is to define extensions to the SQL language itself \cite{DBLP:dblp_conf/edbt/DobosSBBCTMTJ11, DBLP:dblp_conf/ssdbm/DobosBLSC12}. Extended syntax SQL can be parsed on client side and custom additions be replaced by UDF calls.

One big shortcoming of traditional RDBMS is their lack of horizontal scalability which makes it hard to adapt them to the ever growing data sizes. The complex, multi-layered infrastructure of the existing RDBMS software makes implementation of transparent scale-out techniques into the original code base prohibitive and database cluster software built around the major products tend to implement distributed functionality as a higher level wrapper using client API.

\subsection{The Graywulf Project}

Our long-term goal is to build software that will convert a bunch of database servers into an end-to-end scientific data warehouse solution. Functionality of such system should include
\begin{itemize}
\setlength{\parskip}{0pt}
\setlength{\parsep}{0pt}
\item integrated scientific data analysis inside the database,
\item transparent distribution of computations and partitioning of data over a cluster of servers,
\item wide support for data ingestion, data transformations, sharing and publishing of scientific data, and access to remote data sources,
\item support of data types essential for scientific data,
\item support for spatial extensions,
\item extensibility framework to support new data formats, math libraries, query language syntax, etc., and
\item programmable and interactive user interfaces to query, analyze and visualize data,
\item wide support for metadata and provenance information, and
\item cluster management and monitoring.
\end{itemize}
As we will discuss later in details, a significant fraction of these features is already in place, certain features already exist as stand-alone software but are not integrated into the Graywulf system, while other components are still in the phase of discussion and design. Graywulf is built around Microsoft SQL Server. Our decade-long experience with it and its easy extensibility via .Net~runtime integration made it an obvious choice for the project. Most of the code is written in \csharp and runs on .Net~4.0 and depends on .Net Workflow Foundation.

\section{RDBMS for science}
\label{sec:rdbms}

As it was mentioned in the introduction, in order to successfully use RDBMS for the purpose of scientific data analysis, several extensions had to be developed. We briefly review some of the existing extensions and discuss future work in this section.

\subsection{Federated databases}

The performance of database servers depends largely on the I/O subsystem. Reasonable performance can only be achieved by using finely-tuned, high-speed directly attached storage which (in case of commodity servers, as of 2013) is limited somewhere around the 100~TB range in size per server node. This makes it necessary to think about the possibilities of scaling-out databases to multiple servers. Efficient database sharding, while a basic feature of most no-SQL systems, is not part of most mainstream RDBMS products yet.

High performance data warehouses might require setting up multiple servers for the same data set in order to serve the high number of requests. While difficult in case of OLTP systems, load balancing over a set of identical databases is not a particularly complex problem in case of the mostly read-only scientific data warehouses. Load balancing queries in a round-robin fashion is really simple for lots of cheap queries, a single long-running, expensive query, on the other hand, might require partitioning the problem (for example by primary key ranges) and executing it on multiple servers in parallel.

Scientific data is often stored in a heterogeneous distributed environment. Even if we can stick to a single platform for the core of the data warehouse, as we do in case of Graywulf with Microsoft SQL Server, remote data on different platforms should be transparently accessible. This rises several problems including different transfer protocols, user authentication, data formats, SQL flavours, caching of remote data etc. We will get back to some of these issues in Section~\ref{sec:caching}~and~\ref{sec:transformations}.

\subsection{Multi-dimensional databases}
\label{sec:array}

Scientific data are most often multi-dimensional. The spatial and temporal coordinates and the measured quantities can be considered as data points of a large parameter space. If the coordinates are discrete and quantities are measured at all coordinate points we usually talk about \textit{grid data}. It has to be clearly distinguished from the case when coordinates are continuous, the parameter space is sampled sparsely but the number of data points is high. We will call this latter type \textit{point clouds}.

There is a huge effort in the scientific database community to introduce arrays into databases as first class citizens \cite{DBLP:dblp_conf/edbt/KerstenZIN11, DBLP:dblp_conf/ssdbm/StonebrakerBPR11, DBLP:dblp_journals/debu/IdreosGNMMK12}. Some products, like \textit{rasdaman} \cite{DBLP:dblp_conf/sigmod/BaumannDFRW98}, primarily target geosciences and implement array support in the form of client libraries to a wide selection of mainstream database server products. These efforts concentrate on storing a moderate number of huge, densely sampled grid data sets like images or data cubes.

Other databases consist of a large amount of small arrays (typically in the 100 byte to kilobyte range). The point clouds require a significantly different storage model and an entirely different set of indexing and search algorithms \cite{2007AN....328..852C}.

\subsection{Point cloud databases}

The simplest spatial indexing techniques divide the (usually finite) surface or space into disjoint cells and data points are tagged with the cell's identifier that they fall into. By knowing the cells boundaries, their volumes and the number of contained points a bunch of questions can be answered about the data points only by accessing the cell information and without loading the data points themselves into memory. The actual data points are only accessed when absolutely necessary. By choosing the ratio of the number of cells and data points right a significant speed-up of spatial queries can be achieved when compared to full scans. For instance, finding all data points within a given sphere goes as follows. Find all cells that are inside the sphere and all those that intersect with it. Data points belonging to cells entirely contained by the query sphere will surely be part of the result set, whereas points within cells intersecting with the query region will have to be examined one by one.

When point clouds are stored in the relational model, a typical fact table containing the data points consists of an ID column that identifies the data point uniquely, a CellID column that stores a reference to the cell which encompasses the point, a vector Coords for the coordinates and another vector Data for storing the measurements. Because fast spatial access is the objective, the primary key is composed of the CellID and ID columns, ordered by the CellID first. This storage model allows for loading data points of a single cell sequentially.

Geographic information systems (GIS) have long been using 3D and spherical indexing techniques to achieve high performance at spatial queries. GIS implementations, however are not open enough to support the special needs of astronomy, for instance, where many problems require large-scale clustering analysis of the spatial distribution of hundreds of millions of data points. To support the requirements of astronomy, Budavari et al. developed a Spherical Library for Microsoft SQL Server that enables astronomers to describe spherical regions (areas on the sky bounded by great circle arcs) analytically and perform boolean operations with them. The Spherical Library contains an implementation of the Hierarchical Triangular Mesh (HTM) indexing \cite{2007ASPC..376..559B, 2010PASP..122.1375B}. Lemson and Budavari developed a simple library to index data distributed in the euclidean 3D space that also runs inside the database server \cite{DBLP:dblp_conf/ssdbm/LemsonBS11}. Csabai et al. designed similar software tools for higher dimensional metric spaces and non-uniform point distributions \cite{2007AN....328..852C}.

\subsection{Arrays in RDBMS}

One of the obvious limitation of the type system of RDBMS is the lack of a flexible array type, though certain products do support arrays \cite{PostgreSQLArrays}. Mainstream RDBMS products have been shown to be successfully used to store and process point cloud array data efficiently \cite{DBLP:dblp_conf/edbt/DobosSBBCTMTJ11, 2007AN....328..852C}. The Array Library developed for Microsoft SQL Server by Dobos et al., is optimized for small arrays stored in-page in table columns. In the Array Library array manipulation is done via user-defined function calls. Because of the high number of element types and operations, and the absence of function overloading in SQL, array queries become increasingly complex because of the numerous long and very similar function names \cite{2012ASPC..461..323D, DBLP:dblp_conf/ssdbm/DobosBLSC12}.

Efforts to design a query language especially for arrays concentrate on grid data and treat arrays at the same level as table. In case of arrays as column data types a syntax like that of PostgreSQL is more favorable  \cite{PostgreSQLArrays}.

\subsection{SQL Language extensions}
\label{sec:syntax}

Implementing spatial extensions that run inside the database process or adding array support to relational databases is a task for programmers but the libraries will be used by scientists who are only familiar with the SQL language. Consequently, extensions to SQL are necessary to support array and spatial notations. One of the key goals of Graywulf is to build a library that allows for simple implementation of SQL extensions, as we will explain it in Section~\ref{sec:parser}.

\section{Anatomy of SQL queries}
\label{sec:query}

The central part of Graywulf is the distributed query engine that can calculate table joins across a federation of databases. Our goal is to enable scientists to access all distributed data transparently via simple SQL queries and hide all the complexities mentioned above. The problem of optimizing and executing distributed join queries, in general, is rather hard. There are, however, certain simple scenarios which can be easily implemented and which make access to distributed data much simpler. 

If we can make the assumption that most of the data required to execute a query is available co-located on one of the data warehouse servers, it seems feasible to fetch all other data from remote servers and other various data sources, copy these data into local tables, and finally execute the join query locally. To perform such an operation the execution environment must find the server containing most of the required data, copy remote data there and execute the final join.

\subsection{Distributed join queries}
\label{sec:distributed}

Graywulf supports joins between co-located tables and remote data sources. Before executing the joins, remote tables are copied to the database server containing the co-located databases. This is obviously the simplest solution to the distributed join problem and there is a lot of room for optimization. The current version of the join engine is able to find the server with the most co-located data to execute the join. To minimize the amount of data to be copied from remote servers, the engine can also restrict queries fetching remote tables to the necessary columns and figure out, based solely on the query text, the most restrictive constrains that can be imposed on the remote tables without losing the ability to correctly calculate the results of the join operation. This is done by analyzing the WHERE criteria.

\begin{figure*}[t]
\includegraphics[width=\textwidth]{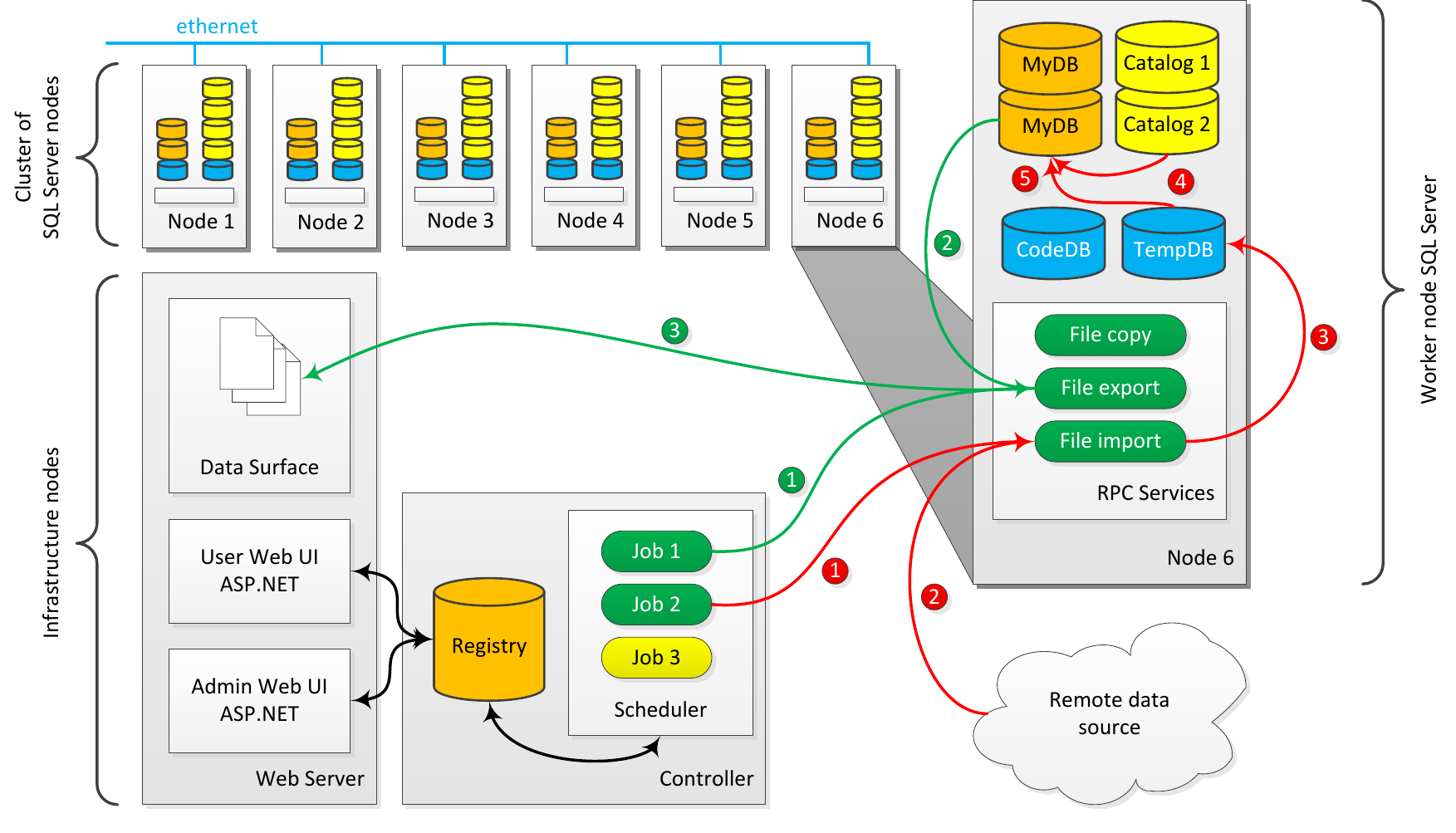}
\caption{Physical system architecture of a typical Graywulf cluster. The basic configuration consist of a controller machine, a web server for the user interfaces and any number of worker nodes. Graywulf registry and the scheduler are co-located on the controller for performance reasons. Worker nodes usually contain data archives, MyDBs, TempDBs for temporary data and CodeDBs for user-defined functions and stored procedures. For the description of arrays and number, please see Section~\ref{sec:distributed}.}
\label{fig:query}
\end{figure*}

Fig.~\ref{fig:query} explains the steps of a distributed join operation (red arrows and numbers). When a user submits a distributes join query it becomes a job and gets into the job queue. Eventually, the scheduler service picks the job up, identifies it as a query job and starts its processing. The query is parsed and all necessary databases and tables are identified. Based on the Graywulf registry (c.f.~Section~\ref{sec:cluster}) the scheduler is able to figure out which cluster node contains most of the data co-located and assigns the processing to that server. First of all, remote tables have to be copied to the node. Because remote data sources can contain data in any format, an import task is scheduled on the worker node (Step~1). The import operation consist of the execution of a query against the remote data source (Step~2). This step requires translating certain parts of the original query to the particular SQL accent the remote data source speaks. In Step~3 remote data is fetched and bulk-inserted into a temporary database (TempDB) on the worked node. Multiple remote tables are fetched in parallel. If the remote data source supports it, indices created on the remote tables can also be created on the cached versions of the tables in TempDB, which helps perform the joins significantly. Once all remote tables are copied to TempDB, the join between the local tables and the cached remote tables can be executed (Step~4). The results of the join get stored in the MyDB of the user (Step~5).

To reduce the overhead of remote table caching an algorithm figures out the combination of table constraints that minimizes the number of rows to be copied from the remote database. The algorithm works solely from the query, no table statistics are used. Still, it is superior to the remote table join algorithm implemented in Microsoft SQL Server which either retrieves entire tables, or queries records one by one.

\subsection{Partitioned queries}

To distribute a single query over a set of server containing replicas of the same data we implemented an experimental partitioned query engine. A partitioning column can be defined on the first table following the FROM clause and the system is able to automatically figure out the partition boundaries. First a histogram of the distribution of values in the partitioning column is calculated. When the histogram is calculated, all constraints of the original query are imposed on the table.

In general, histogram calculation is a full scan operation which can take a significant time if executed on large tables or indices. Since it would mean an unacceptable overhead, we create a small, randomly subset of each data set registered in the framework and run value statistics queries on these small databases. The random sampling rate is usually 0.1\% and only the main fact table is sampled. When sampling other tables, we simply enforce foreign-key constraints.

\subsection{Parsing queries}
\label{sec:parser}

The first step of distributed query processing is to build a parsing tree from the query. Graywulf uses its own SQL parser to perform this task. As a second step to query parsing, identifiers have to be resolved against the underlying database schemas. Though name resolution is normally done by the database server, to handle remote data sources (for example web services that return data tables in custom formats) a specialized name resolver had to be written. During name resolution all tables and columns are identified that are referenced by the query and required to execute the joins. By having the list of necessary column, queries for caching remote tables can be easily generated.

Name resolution is supported by a schema handler layer which offers a uniform view of database schemas regardless of the underlying software product. Schemas are extensively cached to minimize the number of schema queries against the database servers.

To create a SQL parser we implemented a parser generator from scratch that suits our specific needs and generates recursive descent parsers. Parsing tree nodes are generated as partial \csharp classes, so auto-generated code and hand-written code can be kept separately. 

The internal representation of the queries in the Graywulf libraries is always the parsing tree. Whenever a transformation to the query is required (for example removing the ORDER~BY clause, replacing array notation with direct function calls, or substituting remote table references with locally cached table names, etc.) it is made by modifying the parsing tree and then rendering it to text to send it to the database server. When rendering code from the parsing tree certain rules can be overriden to support various flavors of SQL. To hide the complexity of the parsing tree from the other parts of the API, the parser library implements lots of functions to simplify parsing tree traversal.



To make extending the SQL syntax simple, our parser generator supports grammar inheritance. The Graywulf query processing API is designed to use multiple SQL syntaxes, and select the right parser. Additional parsers and name resolvers can be added to the system as plug-ins.

\subsection{Remote table caching}
\label{sec:caching}

Executing distributed joins in Graywulf is very simple and will be improved in the near future. All remote tables are copied to the \textit{TempDB} of a single server containing most co-located data.  Queries are then rewritten to work from the locally cached version of remote tables, and executed directly by a single RDBMS instance. In the current version of Graywulf tables are copied from the remote data source every time a query is executed. Obviously there is a lot of room for optimizations here. When working on a typical data analysis problem, scientists tend to execute the very same queries over and over to tweak certain parameters. If remote data is needed to process queries caching them locally once and reusing them from the local copy is an evident way to speed up operations significantly.

Li in his PhD thesis \cite{NolanPhd} analyzed the problem of efficient caching of remote tables thoroughly and implemented a prototype system called TileDB. To reduce network traffic several optimization steps can be taken. The most obvious is to only fetch that part of the remote data that is absolutely necessary to evaluate a query. It is relatively easy to find the list of required columns and figure out the most restrictive WHERE clause that can be applied to the remote table before fetching the data. There is, however, a trade off between the cache hit rate and the fraction of columns and rows cached. If only a tiny fraction of the table is cached it is very likely that future queries will miss the cache. Li used column usage statistics extracted from a huge set of historical queries to maximize the cache hit rate by adding more columns and pruning WHERE criteria when retrieving tables from the remote databases.

\subsection{MyDB and shared scratch space}

By default results of queries are always stored in the users' MyDB. This becomes a significant bottleneck in situations when multiple queries are necessary to process a data set as intermediate steps might produce data orders of magnitude larger than typical MyDB sizes. Furthermore, MyDBs are usually not on the servers where the source databases reside and copying huge result sets would cause high network traffic and slow down the entire system significantly. In a future version of Graywulf we will make it possible for users to use shared space on the worker nodes to store intermediate query results on the worker nodes, next to the source databases. Tables stored in this scratch space will have a limited life time but will not have the strict space limits as MyDBs. Because one of the main objectives of Graywulf is to hide the complexity of database clusters from the users, referencing the temporary tables from SQL queries should be simple and should not require any knowledge about the configuration of the cluster. This might rise many optimization problems, for instance, once an intermediate table is created on a worked node, all queries referencing the that table should be routed to the same server.

\section{The scientific data exchange}
\label{sec:transformations}

One of our mid-term goals with Graywulf is to make it an extensible data exchange for scientific data. One of the cornerstones of providing a uniform, transparent, SQL-based access to all data is to incorporate remote data sources and user data into the same framework. Users will not only be able to access information already in the data warehouse but also to upload their own data into their MyDBs, or to register and query remote data sets. Data coming from outer sources will be automatically converted into an internal representation and inserted into database tables or array structures. Extensions will be in place for accessing arrays and other non-relational structures to make all data queryable from SQL. 

\subsection{Data models and formats}
\label{sec:datamodel}

Scientific data can be very complex, still a handful of relatively simple data models are usually enough to store them. Tabular, object oriented, grid array, hierarchical and graph are the most fundamental data models. Relational databases are limited to store data in tables while some ongoing efforts will very soon make grid arrays available in some database systems. Two other common models, hierarchical and graph data, can be easily mapped to the relational model, though query performances might not be optimal. Scientific data is usually a combination of the aforementioned fundamental data models. For instance, measurements stored in a grid array are usually accompanied by complex metadata stored in hierarchical structures. A good example is the Characterization Data Model of the astronomical Virtual Observatory which defines a standard for tagging grid array data with metadata \cite{CharacterizationDB}.

Data models is only one side of the problem, disk and wire data representation being the other. Every field of science is abundant in data formats which usually have different dialects from research group to research group. This problem is well-known in reference to business applications, and to address it standards, like the Electronic Data Interchange (EDI), were developed. Still, many applications have their own representation of the same logical data models and communication between them requires a software layer that can translate between the representations. Several server applications (Microsoft BizTalk Server, IBM WebSphere, Oracle SOA Suite, etc.) with numerous adaptors to support various standards and data representations exist to help integrate business systems.

We have designed and partially implemented a plug-in based system that will make writing data format adapters simple. The current version of Graywulf focuses on table data only, and data formatter plug-ins must implement ADO.NET interfaces to return the table schema and iterate through the table rows sequentially. Via these ADO.NET interfaces formatters are tightly integrated with the database server to achieve the highest possible bulk-insert performance. Serialization of tables and query result sets is done a similar way, via formatter plug-ins.

\subsection{Data surfaces}
\label{sec:datasurface}

A data exchange system must have interfaces to ingest and retrieve data via various network protocols and file or wire formats. Some of the most common data surfaces use the SOAP or REST standards or simply return data as downloadable files via HTTP. Higher throughput systems, however, might require more sophisticated wire protocols in the future. The current version of the Graywulf code supports only uploading and downloading tables in CSV files, but most of the formatter plug-in system is in place.

In general, data stores might or might not be aware of the data formats they store. In an ideal system data file formats would be automatically detected as files are uploaded, and converted into an internal exchange format, thus becoming accessible for data processing services connected to the data store. In our case, for example, if tabular data were uploaded via the data surface, it would readily be materialized as a table in the user's MyDB, be queryable using SQL, and be downloadable in any of supported file formats.

\subsection{Metadata and provenance}
\label{sec:metadata}

It is very important to distinguish two types of metadata. The first type is what the natural scientist considers important: data describing the circumstances of measurements. These could be the geographic and temporal coordinates, environmental parameters, instrument properties and setting, etc. The second type of metadata is what the computer scientist considers important: information about the content of the variables. In this section we concentrate on the latter type of metadata. While the structure of the first type of metadata is specific to the field of science, the second type is rather generic and can be implemented as a simple extension to the database schema. To clarify the distinction between the two types, let us consider the following example. A database table containing start and stop times for measurements is usually considered a metadata table, while the column names are also, obviously, metadata describing what quantities are stored in the columns. Column names themselves, however, are not enough to fully describe quantities. If we want a system that can automatically identify columns storing a given physical quantity (for example find the columns storing celestial coordinates in a given coordinate system automatically) we need, at least, metadata to indicate the type of the coordinate system (equatorial, galactic, etc.), the quantity (which spherical coordinate of the two) and the unit the values are stored in (degrees, radians, sexagesimal notation).

Graywulf contains a framework for handling schema information and additional metadata tightly integrated. Schema objects are tagged with three metadata values: content identifier, unit and human readable description. Content identifiers and unit formats are specific to the field of science and they are stored as text.

Many scientific data sets contain tables with hundreds of columns which makes assigning metadata to already existing schemas a laborious task that usually requires a person with full knowledge of the dataset and some database background. Consequently, the most appropriate time to assign metadata is when the database schema is being designed. Following the idea of Gray \& Szalay \cite{2008CSE....10...38S, 2008CSE....10...30T} Graywulf offers tools to add metadata information to SQL scripts, very similar to XML comments of \csharp in the .Net framework. Listing~\ref{que:sample} shows and example of a SQL script with XML comments. As a first step, a file scanner extracts XML comments from the SQL script. Because XML comments do not contain object names, rather their position inside the SQL script determines what object they refer to, the SQL script has to be processed as well. Once XML comments are associated with object names a standard XML file is written that contains all metadata. After the SQL script is executed and the database is physically create metadata can be added to the database schema by running another tool on the metadata XML file.

\begin{query*}
\begin{minipage}{\textwidth}
\begin{lstlisting}
		CREATE TABLE Star
		(
		-------------------------------------------------------------------------------
		--/ <summary>The objects classified as stars</summary>
		--/ <remarks>Contains photometric parameters of point-like objects,
		--/ including quasars</remarks>
		-------------------------------------------------------------------------------
			ObjID bigint,  --/ <column ucd="id.main">Unique SDSS identifier</column>
			ra    real,    --/ <column ucd="pos.ra" unit="deg">Right ascension</column>
			dec   real,    --/ <column ucd="pos.dec" unit="dec">Declination</column>
			...
		)
\end{lstlisting}
\caption[labelfont=bf, textfont=bf]{Sample script demonstrating how metadata is interleaved with SQL code. Lines containing XML comments must be marked with -\--/. Graywulf's metadata tool can extract these comments from SQL scripts and store them in the database schema. See Section~\ref{sec:metadata} for details.}
\label{que:sample}
\end{minipage}
\end{query*}

To store metadata inside Microsoft SQL Server database schemas we use extended properties. In SQL~Server any number of extended properties can be assigned to every schema object down to the table column and function parameter level. Querying extended properties, consequently the metadata, is supported via system views.

We mentioned above that content identifiers and units are usually specific to the field of science. In the near future, however, it will be useful to standardize them. The reason behind this is to be able to build generic semantic tools that will be able to automatically discover relations among data sets and help build complex scientific queries.

Conclusions drawn from data analysis are just as reliable as the data are. As scientific data sharing is becoming more common, it is getting more important to have detailed information about the origin, preprocessing and quality of the data. The fact that in Graywulf all data are manipulated with SQL queries, and we parse those queries before they are executed, makes it possible to follow data provenance automatically.

\section{Server cluster management}
\label{sec:cluster}

To automate system management we created a central registry that stores information about the configuration and status of the entire database server cluster. The registry contains details about the system at the granularity that is necessary to optimize database configurations for the underlying hardware. Although there are lots of products available for server cluster management, none of them provide the functionality to micro-manage a database server cluster at such level we require. Consequently, we ended up implementing a simple management suite ourselves, specifically for Graywulf.

The central registry consist of a hierarchy of objects, each object referring to hardware or software components. Objects are organized into five groups: cluster, federation, layout, jobs and security. The cluster group contains the description of the components of the underlying hardware and software. Objects in the federation group describe the logical organization of data while objects in the layout group store how and especially where data are actually stored. The Graywulf registry is stored in a Microsoft SQL Server database which provides transactional access to the cluster configuration.

\begin{figure*}[t]
\begin{minipage}{\textwidth}
\includegraphics[width=\textwidth]{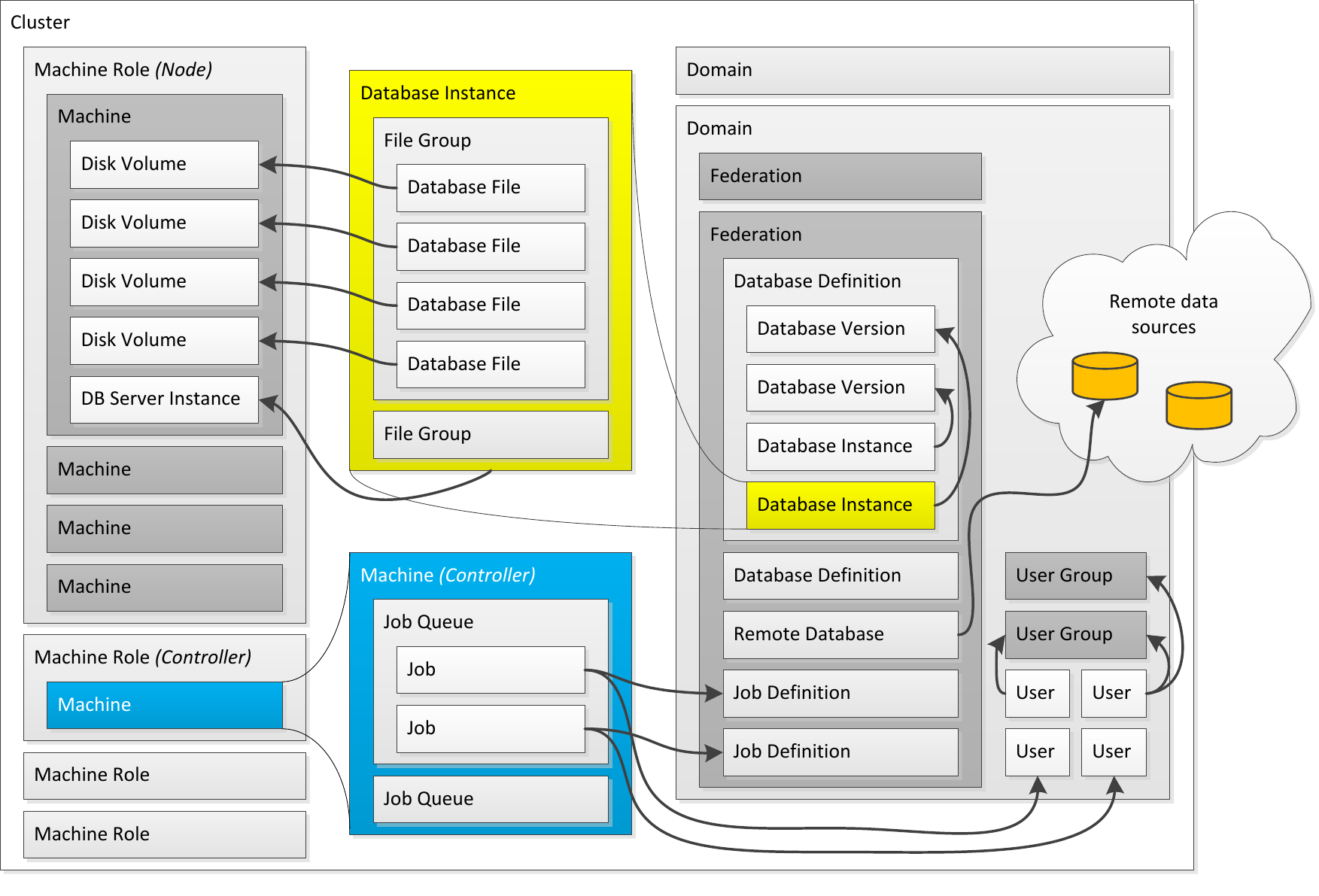}
\caption{Organization of the Graywulf registry. The description of the physical cluster is on the left and logical configuration is on the right. Database instance objects make the connections between the logical and physical layouts of the cluster. See Section~\ref{sec:cluster} for detailed description of the various objects.}
\label{fig:registry}
\end{minipage}
\end{figure*}

Although most details about the servers and databases could be figured out directly from the database server and operating system configurations, for performance and portability reason, we replicate all important information in the Graywulf registry. To help enter information into the registry, and maintain its consistency with the server configurations, the Graywulf library supports querying server settings, comparing them to the actual state of the registry and making the updates, if necessary. Figure~\ref{fig:registry} shows the organization of the objects in the Graywulf registry.

\subsection{Cluster hardware}

The registry stores detailed information about the hardware and software configurations. Machines are organized into machine roles. The two default roles are \textit{controller} and \textit{node}. The controller is the machine storing the registry and running the job scheduler, right now only one controller per cluster is supported but we will have plans to make fail-over sets available soon. Nodes store science data and execute calculations. Further machine roles can be created to differentiate the purpose of servers.

A typical high performance database server may contain two or more RAID controllers, each configured to handle multiple RAID volumes with different redundancy levels for temporary and catalog data. In order to achieve the best database performance on a given hardware, databases have to be configured specifically for the I/O subsystem of the server. In case of Microsoft SQL Server, database files have to be allocated such a way that data get evenly distributed among RAID volumes. Consequently, the registry contains information about the servers down to the disk volume level. Each disk volume has flags to indicate whether it is intended to store the system, transaction logs, temporary data, or science data.

\subsection{Domains, federations and databases}
\label{sec:federation}

Since database server clusters are generally used for multiple projects simultaneously, Graywulf divides the configuration into domains. A domain is a loosely coupled set of data and services that belong to a single field of science. Federations, on the other hand, are collections of databases backing a single service. For example, a domain could be created for astronomical data with two federations: one federation for reduced astronomical catalogs and another for spectra. Each federation would consist of multiple data sets.

Federations contain database definitions (which in turn contain many physical databases) and remote database connections. Database definitions are basically prototypes of the actual physical databases. A database definition consists of an abstract schema but no data. When a database instance is created physically on one of the nodes, it is automatically configured based on the settings of the database definition and the settings of server on which it is instantiated. This is the point when the optimizations for the underlying hardware can be done. All database instances of a database definition should have the same schema but may contain different data. Database definitions can be instantiated as individual databases, mirrored sets or sharded sets (not yet implemented). Each database can have multiple versions. For example, a version can be created that contains all the data while a mini version containing only a small percent of the data can be used to gather query statistics. Remote database connections are references to external data sources that are not managed via the Graywulf system. We currently support Microsoft SQL Server and MySQL as external data sources.


\subsection{Job framework registry}

All the job framework related information is stored in the Graywulf registry. Job queues can be defined on the cluster, domain and federation level. Each queue is associated with a machine role or a specific machine. By default three queues are created, each one associated with the controller machine: for maintenance jobs, for quick jobs and for long-running jobs. 

In Graywulf jobs are created based on job definitions. Job definitions are job prototypes that define the workflow logic behind a job, whereas job instances only define the input parameters of the workflows. For example, a typical job definition is that of a SQL query. It contains all the logic to execute the query but not the query text itself. When a user submits a query a job instance is created with all the necessary parameters set and enqueued in an appropriate job queue to be picked up by the job scheduler. For more details about the job system, refer to Section~\ref{sec:jobs}.

\subsection{Security system}

The security system of Graywulf is rather simple and only provides authentication of users. Users are defined on the domain level, i.e. user registrations are shared among federations and services belonging to the same data set. User can be members of any number of user groups. Authorization has to be implemented in the services built on top of the Graywulf API. This is one big limitation that we want to address in a future version. We are also planning to add support to various open authentication standards like OpenID and OAuth.

\subsection{Logging}

Because all jobs in Graywulf are workflows (see Section~\ref{sec:jobs}), the logging framework had to be designed to work with .Net Workflow Foundation. Fortunately, WF has its own log even routing infrastructure and features sophisticated event filtering which we could highly benefit from. At highest verbosity the contains information about all registry modifications and workflow activity transitions. Exceptions occurring during job execution are logged with full stack trace for easy debugging. Logs are written into a SQL Server database for easy browsing.

\subsection{Avoiding the configuration hell}

Administration of a Graywulf cluster is done via a web-based user interface. Distributed computer systems easily get very complex to configure, especially when the learning curve of software used is steep. Graywulf has a series of features to avoid the configuration hell. The administration interface uses configuration templates to register new system objects (computers, databases, etc.) in the framework. When new objects are added they are automatically preconfigured to the most typical settings. The number of settings is kept minimal which makes the system less flexible but simplifies configuration tasks significantly. Registry objects, once created, can be updated to reflect the actual hardware and database configurations with a single click. This feature, called automatic discovery, turned out to be especially useful when working with already existing database server clusters on which the Graywulf system is to be installed. To make a configuration portable any branch of the registry hierarchy, including the entire configuration, can be serialized into XML. XML files can later be merged into an existing configuration. 

\section{Job system}
\label{sec:jobs}

Data warehouses must support batch processing. We set the following requirements for the batch system.

\begin{itemize}
\setlength{\parskip}{0pt}
\setlength{\parsep}{0pt}

\item Jobs are very complex, parallel workflows.

\item Allow build workflows dynamically at run time (for example based on query plans).

\item Allow queuing of jobs. Queues can process a given number of outstanding jobs at a time and can have different time out periods.

\item Activity scheduler must be data co-location ever. Take computation to the data and not data to the processing.

\item Support suspension, persistence and restarting of long-running jobs. Suspending jobs might be necessary when the system is reconfigured, for example databases are replaced with new versions.

\item Support partial fail and retry of certain branches of the workflows. Retry might be necessary when a certain component of the system fails but the execution branch can be reconfigured to another database or machine. This is an important feature in case of partitioned queries.

\item Support cancelling of long running jobs. This is not a simple task when multiple branches of a workflow can run in parallel.

\item Enforce time limits on entire jobs. Setting a time out period on individual activities is not enough.

\item Provide detailed logging with configurable granularity to help debugging.

\item Hide complexities of parallel coding with a simple programming model.

\item Support delegating certain activities of a workflow to machines different from the one running the workflow system.

\end{itemize}

\subsection{Every job is a workflow}

In the Graywulf batch system each job is a workflow. Workflow instances are created from prototypes called job definitions. The current implementation supports precompiled job definitions only that can be parametrized, but their logic cannot be modified. Future versions of Graywulf will be able to execute dynamically compiled workflows. Jobs are stored in the registry, ordered in queues. A scheduler process runs on the controller machine that polls the registry for enqueued jobs and executes them in-process. Multiple workflows and, within each workflow, multiple activities can run concurrently. Although workflow execution is not distributed, individual activities can delegate work to remote machines. The simplest way of delegating computation is to execute SQL queries on database servers other than the controller machine. A more sophisticated way of task delegation is also part of Graywulf, see Section~\ref{sec:delegation}.

Most Graywulf jobs are queries but data import and export jobs, bulk file copies, plot generation etc. are to be scheduled as well. Jobs can get very complex because distributed queries are translated into a series of smaller queries that need to be executed in the right order but often in parallel. We chose to implement or batch system on the basis of .Net Workflow Foundation (WF). WF enables assembling complex workflows from activities while hides the complexities of parallel coding and logging. WF also support suspending, persisting and resuming long-running jobs (although only between two activities).

\subsection{Scheduler}

Workflow Foundation itself cannot schedule jobs. WF uses a thread pool to execute activities in parallel, but all other scheduling has to be done by a custom-written component, in our case the Graywulf scheduler. The scheduler basically does two things. It is a queue system that polls the registry for new jobs and job cancel requests, start new jobs, cancels running jobs if necessary, and enforces job time-out periods. Its other task is to be aware of the status of the system and distribute workload over the cluster nodes observing data co-location. When a partitioned job is being executed first it gets partitioned to smaller jobs. Each partition will run entirely on one machine to minimize data movement among cluster nodes. When a partition starts, it is the scheduler that assigns the worker node to the workflow branch based on the availability of the datasets required by the job to complete. For performance considerations the scheduler must run on the same machine where the registry database is located. Currently this machine, the Controller, is a single point of failure in the system. This vulnerability will be addressed in a future version.

\subsection{Task delegation}
\label{sec:delegation}

Because all workflows run on the same machine, the Controller, activities must be able to delegate task to the worker nodes. In case of database queries it is simple, the queries are simply sent to the remote database server and all the processing will happen on the remote machine. Other tasks, such as file import and export, plot generation etc. should also be executed on the worker nodes instead of the Controller. One of our goals was to provide a smooth and transparent programming model for workflow developers so task delegation had to be made simple too. Task delegation is based on Windows Communication Foundation (WCF) which is an integral part of the .Net framework. In WCF, when configured such a way, classes can be instantiated on remote machines. If workflow components are designed with distributed execution in mind, from the programmers perspective instantiation of components in-process or remotely will be totally transparent.

Because the application server designed for the .Net framework (AppFabric) is too heavyweight, we decided to implement our lightweight application server, Graywulf Remote Server. This simple process runs on the worker nodes and waits for requests from the Controller or from other worker nodes. The Graywulf Remote Server is designed to be entirely configuration free, the only requirement is to copy binaries containing the remotely activated components to the worker nodes. Security issues are solved by relying entirely on standard authentication and authorization schemes offered by the Windows Domain and WCF.

\subsection{Workflow components}

To support distributed query execution and several cluster management tasks a bunch of components have been developed. These components all support task delegation (see Section~\ref{sec:delegation}). Components are written for bulk data movement (bulk-insert results of a query executed on another server), delegated file copies among worker nodes, formatted file imports and exports, plotting etc.

\section{User interfaces}
\label{sec:ui}

Graywulf offers web-based user interfaces for administrators and end users. Web interfaces are implemented in ASP.NET using AJAX technology and minimalistic graphics to make the pages more responsive and provide better user experience.

\subsection{Graywulf admin interface}

Administration of a Graywulf cluster is done through a web user interface. The organization of the user interface follows the structure outlined at the beginning of Section~\ref{sec:cluster}. Registry objects can be directly edited using web forms but there are also ``wizards'' for more complex tasks, such as generating database shards or mapping database definitions to the underlying hardware. The administration interface includes a log browser and will contain monitoring tools in future versions. The documentation and reference will also be readily available from the same web site.


\subsection{Query interface}

The main entry point for users in Graywulf is the query interface. Its concept is based largely on CasJobs \cite{DBLP:journals/corr/abs-cs-0502072, 50716924, 6071419}. It consists of a schema browser, a query editor, a job history browser and a MyDB manager. The user interface is directly linked to a domain and a federation (c.f. Section~\ref{sec:federation}), so multiple instances of the user interface can installed for a single Graywulf cluster. The schema browser provides detailed information on all data sets available within the federation, including remote databases. It is based on the same schema library as the SQL name resolver (see Section~\ref{sec:parser}) to benefit from schema caching. The browser also displays metadata and detailed descriptions of the objects. While the current version does not have this feature, plans include adding schema search capabilities to the system. The query editor is a simple page with a text editor that supports SQL syntax highlighting. There are buttons to perform syntax checking and submit queries to various job queues with different time out intervals. We are planning to add auto-completion functionality and a query plan viewer to the query editor in the future. The job history viewer simply lists all past jobs including error information.


\subsection{Visualization}

Fast and simple data visualization is always a key part of a data handling system. Visualization is best done on the server side as spectacular plots often require a lot of data to create. Plotting also has to be scriptable and support file formats used for typesetting. Despite its limitations we chose gnuplot\footnote{http://www.gnuplot.info} as the visualization tool behind Graywulf. To visualize data users can write gnuplot scripts. The syntax was slightly modified to allow executing SQL queries inside gnuplot scripts. When plotting, queries are first executed on MyDB and results are saved into temporary files. The original plotting scripts are modified and queries are substituted with the temporary file names. Output images are then displayed on the web page or made available for download.

\subsection{Integrated shared services}

The query interface is just one of the possibilities of how services can be built on the Graywulf library. Other, not SQL based, but more sophisticated user interfaces can benefit from the shared services Graywulf provides: user authentication, job system and scheduling, MyDB, data transformations etc. The web-form-based user authentication component is designed to be easy to integrate into web sites, just like the error reporting framework which help send detailed error messages and stack traces to the system administrators.

\section{Virtual Observatories}
\label{sec:vo}

The astronomical Virtual Observatory (VO) \cite{DBLP:dblp_journals/fgcs/SzalayB99, 2001Sci...293.2037S} is the main use case of Graywulf and Graywulf's development is mainly driven by its requirements. VO is an excellent example of a heterogeneous, distributed scientific database system. The range of astronomical data sets cover all typical data models we mentioned in Section~\ref{sec:datamodel}. Photos taken of the night sky are grid data, reduced object catalogs fit well into the relational model and cosmological N-body simulations are point clouds. The VO defined its own extensions to the SQL language called Astronomical Data Query Language \cite{ADQL}. ADQL defines a set of extensions to deal with spherical regions, similarly to GIS systems. A good example for a data surface (see Section~\ref{sec:datasurface}) is VOSpace \cite{2007ASPC..382..433G}. VOSpace is a lightweight storage service that uses the REST protocol to store any type or format of data. When users upload their files to VOSpace, the service tries to figure out the data format and, beside saving the original file, saves it in a database to manipulate from SQL. While this is a simple task for tabular data, more complex data models (images, spherical region descriptions) require special libraries to handle. Graywulf was designed to support these kind of operations by its extensible data transformation framework, c.f. Section~\ref{sec:transformations}. The main objective of VO is to federate geographically distributed data set, which is beyond the current focus of the Graywulf project as Graywulf is designed for mostly co-located data sets. One can, however, easily imagine the VO as a federation of multiple Graywulf systems which use standard protocols to share data among each other. The effort to federate astronomical data sets is controlled by the Internation Virtual Observatory Alliance\footnote{http://www.ivoa.net}.

\section{Summary}

\begin{figure*}
\begin{minipage}{\textwidth}
\includegraphics[width=\textwidth]{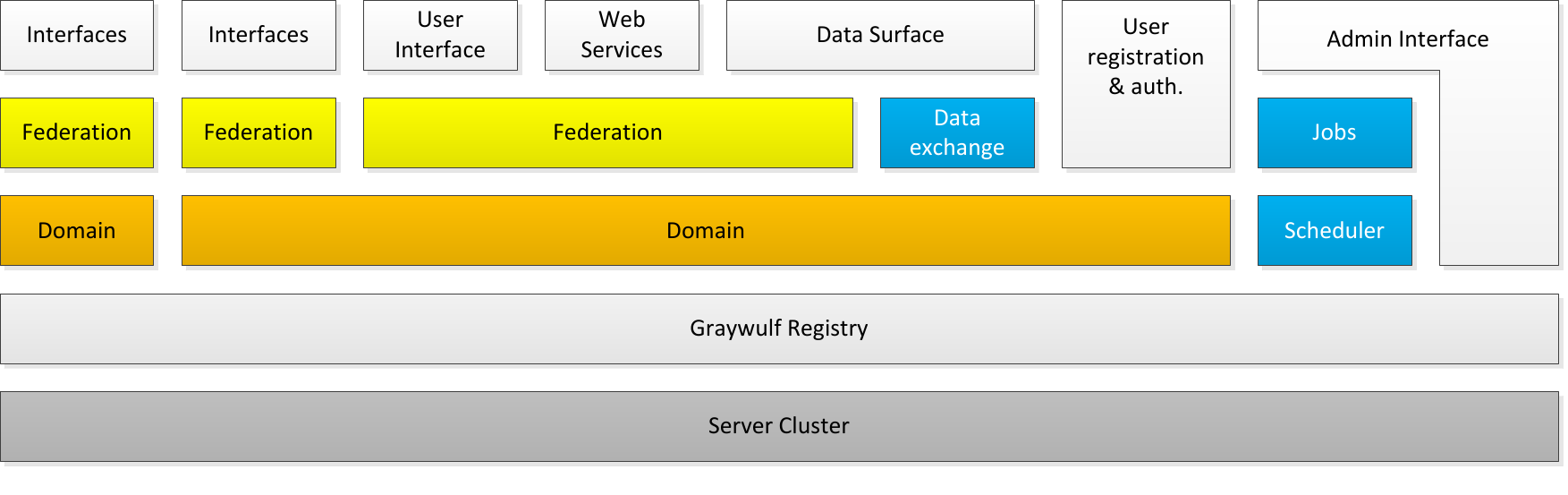}
\caption{Application layers of a Graywulf cluster.}
\label{fig:layers}
\end{minipage}
\end{figure*}

In this paper we have given a detailed introduction to the Graywulf system, an API and runtime developed for distributes data processing using Microsoft SQL Server. Figure~\ref{fig:layers} summarizes the application layers of a Graywulf configuration. While many components of the system are under continuous development, the current version is already operational and acts as the backbone for SkyQuery, the astronomical probabilistic cross-identification service \cite{DBLP:dblp_conf/ssdbm/DobosBLSC12}.

In the future we want to focus our efforts on two development directions. We will convert existing Virtual Observatory services for Graywulf to build an integrated data warehouse as it was outlined in Section~\ref{sec:vo}. The implementation of a more sophisticated distributed query engine is our long-term plan.

\section*{Acknowledgments}
This research is partly funded by the Gordon and Betty Moore Foundation through Grant GBMF\#554.02 to the Johns Hopkins University. The research was also supported by the NSF grant OIA-1124403. This work was partially supported by the European Union and the European Social Fund through project FuturICT.hu (grant no.: TAMOP-4.2.2.C-11/1/KONV-2012-0013). The project was also supported by the Hungarian grants OTKA 103244, NAP 2005/ KCKHA005 and the MAKOG Foundation.

\balance

\bibliographystyle{abbrv}
\bibliography{graywulf_ssdbm}  

\end{document}